\documentclass[a4paper,11pt]{article}
\usepackage{pos}
\usepackage{placeins}
\usepackage{siunitx}
\usepackage{subcaption}
\usepackage{xfrac}
\usepackage{csquotes}

\renewcommand{\logo}{\relax}

\DeclareSIUnit\neq{~\text{n}_{\text{eq}}/\text{cm}^2}

\usepackage[numbers]{natbib}
\title{Observation of Charge Enhancement in forward-biased neutron-irradiated 4H-SiC PiN Detectors in UV-TCT Measurements}
\ShortTitle{UV-TCT Charge Enhancement in forward-biased 4H-SiC Detectors}

\author*{Andreas Gsponer}
\author{Philipp Gaggl}
\author{Jürgen Burin}
\author{Simon Waid}
\author{Thomas Bergauer}

\affiliation{Institute for High Energy Physics, Austrian Academy of Sciences,\\
Nikolsdorfer Gasse 18, Vienna, Austria}
\emailAdd{andreas.gsponer@oeaw.ac.at}

\abstract{
Due to the increased commercial availability, wide-bandgap semiconductors and their radiation hardness have recently received increased interest from the particle physics community.
4H-Silicon Carbide (SiC), especially, is an attractive candidate for future radiation-hard detectors which do not require cooling.

This paper investigates the radiation hardness of 4H-SiC p-in-n detectors irradiated up to \SI{5e15}{\neq} using UV-TCT.
The samples have been operated in reverse and forward bias, which is possible due to the heavily decreased forward current after irradiation.
Previous studies have already hinted at an excessive charge collection in forward bias, even exceeding a charge collection efficiency (CCE) of \SI{100}{\percent}.
In this work, the excessive CCE in forward bias was shown to correlate heavily with the spatial profile of injected charge.
For a sufficiently focused laser, the CCE starts to increase at high forward bias and even surpasses \SI{100}{\percent} instead of saturating as it does for a defocused laser beam.
In reverse bias, the CCE was found to be independent of the beam spot size.
For samples irradiated to high fluences (\( \geq\SI{1e15}{\neq}\)) the excessive CCE in forward bias is smaller and negligible at the highest fluences.

Additionally, the CCE was observed to correlate to the rate of charge injection (laser pulses per second), with a logarithmic increase of the collected charge if a threshold of injected carrier density is exceeded.
The mechanism of these effects is still an ongoing topic of study, however, the observations already pose implications for the accurate experimental characterization of irradiated SiC detectors.
}

\FullConference{6th International Conference on Technology and Instrumentation in Particle Physics (TIPP2023)\\
 4 - 8 Sep 2023\\
Cape Town, Western Cape, South Africa\\}

\begin{document}
\maketitle

\section{Introduction}
The radiation hardness of Silicon Carbide (SiC) detectors has been an ongoing topic of interest since its first applications as a detector for ionizing radiation~\cite{Babcock_1965}.
Due to its wide-bandgap, the leakage current of SiC devices is extremely small (\si{\pico\ampere} or below), even after irradiation and at elevated temperatures~\cite{CNM_SiC_Diode, waid_detector_2023}.
Together with the high charge carrier velocity in 4H-SiC (the polytype most commonly used in industry) and it's commercial availability, this present a promising candidate for future particle physics experiment, which pose significant timing and radiation hardness constraints.

After neutron irradiation of 4H-SiC diodes to fluences above \SI{1e14}{\neq}, the current in forward bias has been observed to drop to very low levels (at the highest fluences even indistinguishable from reverse bias), which allows for operation of the devices as detectors in this regime.
Operating irradiated detectors in forward bias has already been demonstrated for silicon detectors~\cite{Beattie_2000}.
For 4H-SiC, however, an unexpected increase in the charge collection efficiency (CCE) has been reported in UV-TCT studies~\cite{gsponer_neutron_2023, alvarez_TPA_TCT}.
The CCE can even exceed \SI{100}{\percent}, which implies that more charge is collected than originally deposited.

In this paper, we investigate this yet unexplained behavior of irradiated detectors in forward bias using 4H-SiC p-in-n samples irradiated to \SI{1}{\mega\electronvolt} neutron fluences between \num{5e14} and \SI{5e15}{\neq}.
In detail, the charge collection efficiency in forward biased detectors is investigated as a function of the bias voltage and laser beam spot size (Section~\ref{sec:focus}), as well as the laser repetition rate (Section~\ref{sec:rate}).
The laser optics used for this study are characterized in Section~\ref{sec:optics}.

\section{Materials and Methods}
\subsection{4H-SiC Samples}
The samples used in this study are \SI{50}{\micro\meter} thick 4H-SiC p-in-n diodes, manufactured by IMB-CNM-CSIC~\cite{CNM_SiC_Diode2}.
The active area of the sensors is \(\SI{3}{\milli\meter}\times\SI{3}{\milli\meter}\).
More detailed information on the device cross-section and electrical performance can be found in~\cite{gsponer_neutron_2023}.
The samples used in this specific study do not feature any metallization covering the active area (except for the rim of the active area and the bond pads). Consequently, the entire area can be probed by a laser, which eliminates the dependency of the signal on the beam spot size. On the flipside, the detector signals has to be transferred to the readout by the $p^{+}$~implant, which has a smaller conductivity.

Three samples have been irradiated to fluences of \num{5e14}, \num{1e15} and \SI{5e15}{\neq}  at the TRIGA Mark II reactor at the Atominstitut in Vienna~~\cite{gaggl_charge_2022}.
Metallized samples that have been irradiated in the same campaign have already been thoroughly studied using UV-TCT~\cite{gaggl_charge_2022}, $\alpha$-particles~\cite{gaggl_performance_2023} and I-V / C-V plus proton beams~\cite{gsponer_iworid_2023}

\subsection{UV-TCT Setup}
A PILAS PIL1-037-40FC laser was used to generate pulsed \SI{370}{\nano\meter} UV light with a pulse width of less than \SI{50}{\pico\second}.
Different power-per-pulse and repetition rates can be selected by the laser driver, allowing a power-per-pulse up to \SI{6}{\pico\joule} and repetition rates up to \SI{40}{\mega\hertz}.
The laser was fiber-coupled to the laser optics which are mounted onto a motorized Z-stage which can be used to optimize and change the laser focus.
Together with the picosecond-pulsewidth laser pulses, the laser driver also provides a low-jitter ($<\SI{3}{\pico\second}$) trigger, which can be used to average multiple measurements and  significantly decrease the noise floor.
A Thorlabs S150C power meter was used to measure the time-averaged power delivered by the laser.
More information on the setup can be found  in~\cite{gaggl_charge_2022}.
The silicon carbide sensors were mounted on a readout PCB with an external Cividec Cx-L charge-sensitive amplifier featuring an integrated high-voltage bias tee.
The charge-sensitive amplifier has been calibrated using a well-understood, unirradiated 4H-SiC sample together with a $\alpha$-source in vacuum in order to provide readings in \si{\femto\coulomb} of collected charge.
The PCBs were mounted on a motorized XY-table used to probe different regions of the detectors.
A Rohde\&Schwarz RTP164 oscilloscope in the \SI{16}{\bit}-HD mode was used to measure the amplitude of the signals and the high-voltage bias was provided by a Keithley 2470 source-measure unit.

For SiC samples with no metal covering the active area, the resistivity of the implant introduces a \(RC\) circuit, which can significantly impact the shape of obtained transient waveforms when using a high bandwidth readout~\cite{Lopez_Paz_2024}.
This is similar to the operational concept of resistive silicon detectors (RSDs)~\cite{RSD}.
In order to avoid this effect, the charge-sensitive amplifier in this work uses a long shaping time (\SI{1.2}{\micro\second}) ensuring uniform charge collection over the entire sensor area.

\subsection{Laser Optics Characterization}
\label{sec:optics}
\begin{figure}[htp]
    \centering
     \begin{subfigure}[b]{.415\textwidth}
        \includegraphics[width=\textwidth]{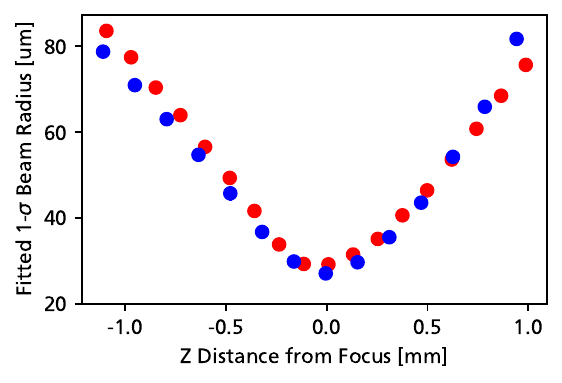}
        \caption{Beam radius fitted from Knife-Edge measurements for the X (blue) and Y (red) axes.}
        \label{fig:laserfocus}
     \end{subfigure}
     \hspace{0.03\textwidth}
    \begin{subfigure}[b]{.44\textwidth}
        \centering
        \includegraphics[width=\textwidth]{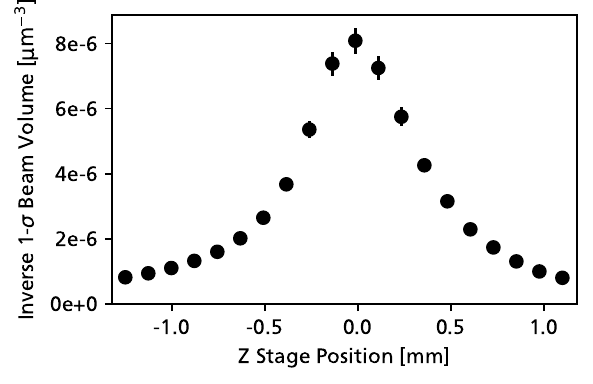}
        \caption{Inverse volume (density) of laser beam inside detector as a function of the focus in Z.}
        \label{fig:powerdensity}
    \end{subfigure}
    \caption{1-$\sigma$ beam diameter and inverse volume (density) as a function of the Z stage position.}
    \label{fig:knife_edge}
\end{figure}

To investigate the dependency of charge collection on the injected charge density, a precise knowledge of the laser optics is required.
Most important is the beam size and divergence of the laser optics, which have been determined using an \textit{knife-edge} scan for both axes (X and Y).
As a sharp-edge feature, the bond pads of the device are used to block out the laser light.
Figure~\ref{fig:laserfocus} shows the resulting 1-$\sigma$ beam radius.
For an optimal focus, a beam waist size of around \SI{28}{\micro\meter} is reached in X and Y.
The volume of the laser beam inside the detector, convoluted by the absorption coefficient of \SI{370}{\nano\meter} UV light in \SI{50}{\micro\meter} 4H-SiC epi ($\alpha = \SI{42.25}{\per\centi\meter}$~\cite{gaggl_charge_2022}), has been calculated as a function of the Z focus in Figure~\ref{fig:powerdensity}.
Between a laser beam which is far away (\(> \SI{1}{\milli\meter}\)) from the focus (\enquote{de-focused}) and the best focus the charge density varies by a factor of around \num{8}.
\section{Results on Irradiated Diodes}
For each sample, the charge collection as a function of the bias voltage has been measured for voltages up \SI{1000}{\volt} in forward and reverse bias.
For the sample irradiated to \SI{5e14}{\neq}, the maximum forward bias used was limited to \SI{270}{\volt}, due to a sharp increase in the leakage current (see~\cite{gsponer_neutron_2023}).
For all measurements, \num{500} waveforms have been acquired and maximum of the waveforms (proportional to the collected charge) has been averaged.
\subsection{Influence of Focus on Collected Charge}
\label{sec:focus}
\begin{figure}[b]
    \centering
    \includegraphics[width=.68\textwidth]{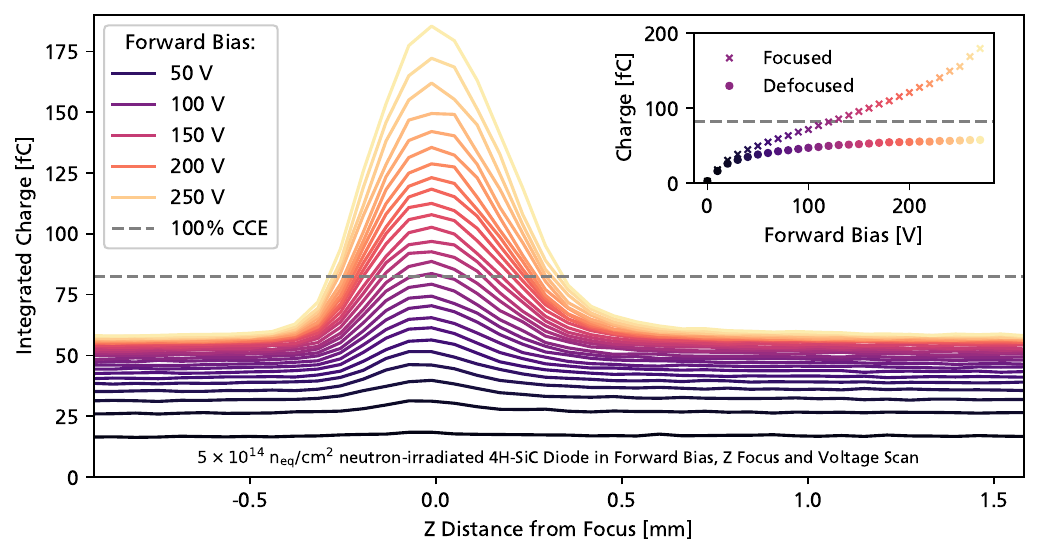}
    \caption{Collected charge in forward bias as a function of laser focus and bias voltage for a laser repetition rate of \SI{600}{\hertz}. \SI{100}{\percent} charge collection efficiency is indicated by a dashed line. At a tight laser focus and for higher bias voltages, \SI{100}{\percent} charge collection efficiency is surpassed. The inset on the top right shows the behavior for a \enquote{defocused} and the a maximally focused beam spot.}
    \label{fig:cce_vs_focus}
\end{figure}

Figure~\ref{fig:cce_vs_focus} shows the collected charge for the SiC sample irradiated to \SI{5e14}{\neq} as a function of the Z-stage position (laser focus) and at different bias voltages.
If the laser is sufficiently de-focused, the charge collection efficiency (CCE) increases as a function of the bias voltage, up to a limit of around \SI{70}{\percent}, where it saturates.
This is consistent with the decrease of the CCE after irradiation due to charge carrier trapping.
\begin{figure}[!!b]
    \centering
    \includegraphics[width=.68\textwidth]{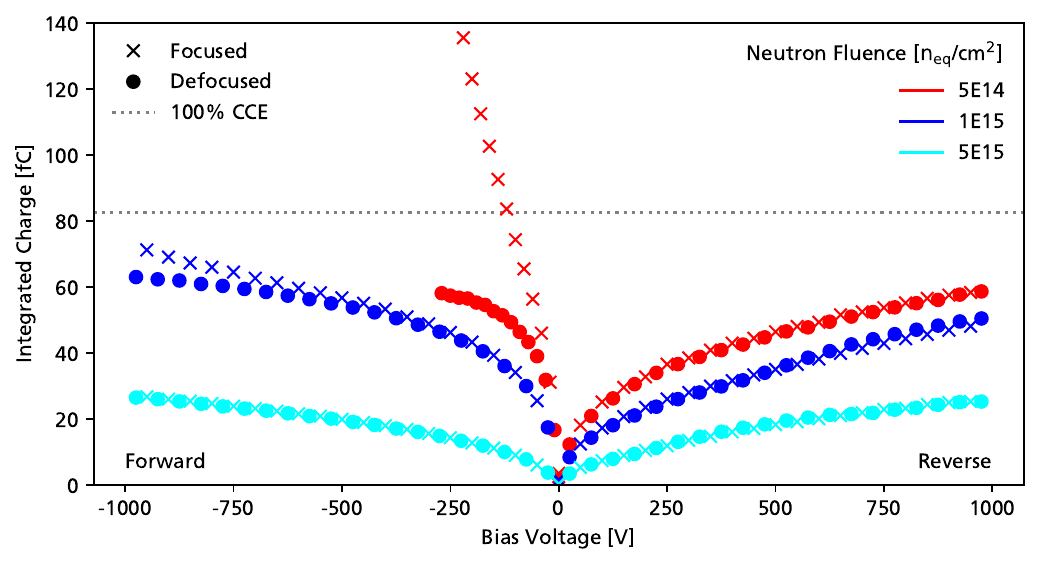}
    \caption{Collected charge for samples neutron irradiated between \num{5e14} and \SI{5e15}{\neq} as a function of the bias voltage. In reverse bias, the charge collection does not depend on the laser focus. In forward bias a clear increase in the signal can be observed for a tightly focused laser.}
    \label{fig:cce_focus_no_focus}
\end{figure}
However, around the position of best focus (highest injected charge density), the CCE shows a pronounced peak, even surpassing \SI{100}{\percent} (relative to before irradiation).
This is a possible explanation for  previously reported measurement results on similar samples~\cite{gsponer_iworid_2023, alvarez_TPA_TCT}

Figure~\ref{fig:cce_focus_no_focus} summarizes this effect together with the same measurements conducted on samples irradiated to \num{1e15} and \SI{5e15}{\neq} as a function of the bias voltage.
Only the extreme cases of the laser focus (focused and defocused) are shown in this plot.
In the reverse bias, the focus of the laser does not have any effect on the collected charge.
In forward bias, the sample irradiated \SI{1e15}{\neq} still shows a increase in the collected charge in the best focus, while as for the most highly irradiated sample (\SI{5e15}{\neq}) no dependency on the laser focus is visible.
\subsection{Influence of Laser Repetition Rate}
\label{sec:rate}
If a highly localized charge injection can influence the local electric field or trapped charge carrier populations leading to an increase in collected charge, then the time-structure of the charge injection also plays a role.
In order to assess the effect of the different injection power densities (\si{ \watt\per\micro\meter^3}), the repetition rate of the laser was changed from \SI{100}{\hertz} to \SI{50}{\kilo\hertz}.
Although the laser controller can operate with repetition rates up to \SI{40}{\mega\hertz} the maximum usable rate was limited by the shaping time of the charge-sensitive amplifier (\SI{1.2}{\micro\second})
Measurements were performed for the sample irradiated up to \SI{5e14}{\neq}, which showed the increase in collected charge the most prominently.
The injected energy density has been calculated by using the volume of the laser beam (measured in the knife-edge scan in Section~\ref{sec:optics}) and the calibration of the charge-sensitive amplifier.
Variations of the laser power per pulse as a function of the repetition rate have been corrected by a reference measurement using a Thorlabs S150C power meter.

\begin{figure}[htp]
    \centering
    \includegraphics[width=.68\textwidth]{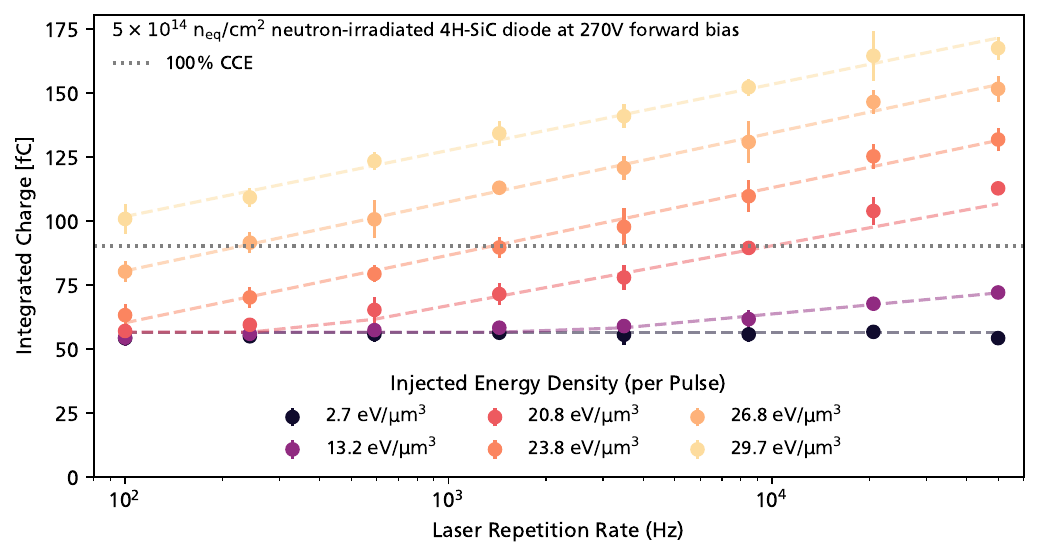}
    \caption{Collected charge for the sample irradiated to \SI{5e14}{\neq} at a forward bias of \SI{270}{\volt} as a function of the injected energy density per pulse (laser focus) and the laser repetition rate. The dashed lines serve as a visual reference.}
    \label{fig:cce_vs_freq}
\end{figure}

Figure~\ref{fig:cce_vs_freq} shows the collected charge as a function of the energy injection density and laser repetition rate.
For the lowest energy density of \SI{1.8}{\electronvolt\per\micro\meter^3} (a defocused laser beam), the collected charge is independent of the laser repetition rate.
At higher energy densities, a threshold can be observed, after which the collected charge increases approximately logarithmically with the repetition rate.
For the highest energy density, an increase in the charge collection can already be observed at repetition rate of only \SI{100}{\hertz}.
\section{Discussion}
The charge collection properties of neutron-irradiated 4H-Silicon Carbide (SiC) diodes have been investigated using a UV-laser TCT setup.
Due to the low leakage currents, the SiC detectors were able to be operated in forward bias and a systematic investigation of the charge collection dependency on factors such as the bias voltage, laser focus and laser repetition rate has been undertaken.

For the sample irradiated to the lowest fluence (\SI{5e14}{\neq}), a large increase in the charge collection was observed in forward bias, when the laser beam spot size was small enough (Section~\ref{sec:focus}).
However, if the laser is sufficiently de-focused, no excessive signal is observed, and the charge collection efficiency (CCE) saturates as a function of the bias voltage.
In reverse bias the results are identical for a highly focused and a defocused laser.
For the samples irradiated to higher fluences the effect is only visible for the sample irradiated to \SI{1e15}{\neq}, for \SI{5e14}{\neq} the effect is not present anymore (see Figure~\ref{fig:cce_focus_no_focus}).
Not only does the increase in collected charge depend on the focus of the laser, but also the laser repetition rate (number of shots per second).
If the injected energy density (as a function of the focus) is high enough, then the collected signal scales approximately logarithmically with the repetition rate of the laser.

The origin of the excessive charge is still a topic of ongoing research.
One hypothesis is the injection of a large amount of highly localized free charge carriers might temporarily reverse the radiation-induced loss of forward current.
The collected signal is then the sum of the injected free charge carriers drifting in the electric field of the detector, plus a a transient, charge carrier injection-modulated forward conduction.
Simulation of radiation damage in SiC are currently ongoing~\cite{pisa_arxiv, austrochip_arxiv} and will be a valuable tool to better understand this effect.

Finally, it needs to be remarked that these effects are a factor that needs to be taken into account when characterizing the radiation hardness of SiC.
In UV-TCT measurements, the repetition rate of the laser should be kept low (< \SI{1}{\kilo\hertz}) and the injected charge should limited, either by decreasing the power-per-pulse or by spreading out the laser beam over a large area.
However, this might be challenging for highly irradiated samples (where the CCE is very small) and for TPA-TCT, which intrinsically features a very localized creation of free charge carriers.
\section*{Acknowledgements}
This project has received funding from the Austrian Research Promotion Agency FFG, grant number 883652. Production and development of the 4H-SiC samples was supported by the Spanish State Research Agency (AEI) and the European Regional Development Fund (ERDF), ref. RTC-2017-6369-3.

\bibliography{biblio}

\end{document}